\documentclass[12pt]{article}
\usepackage{times}
\usepackage{geometry}
\usepackage{comment}
\usepackage{floatflt}
\usepackage[utf8]{inputenc}
\usepackage{enumitem,amssymb}
\usepackage{ragged2e}
\usepackage{multicol}
\usepackage{tabularx}
\usepackage{multirow}
\usepackage{booktabs}
\usepackage{float}
\usepackage[outercaption]{sidecap} 
\usepackage{threeparttable}

\newlist{thematic}{itemize}{8}
\setlist[thematic]{label=$\square$}
\usepackage{pifont}
\usepackage{graphicx}
\usepackage{tabularx}
\usepackage{sidecap}
\usepackage[utf8]{inputenc}
\usepackage[vietnamese,english]{babel}
\usepackage[dvipsnames,svgnames,x11names]{xcolor}
\usepackage{jheppub}   
\usepackage{sectsty}

\definecolor{DarkGreen}{rgb}{0.0, 0.3, 0.0}
\definecolor{purple}{rgb}{0.5, 0.0, 0.5}
\definecolor{red}{rgb}{1, 0.0, 0.0}
\definecolor{green}{rgb}{0, 1.0, 0.0}

\def\3he{$^3{\rm He}$}

\def\lsim{\mathrel{\lower2.5pt\vbox{\lineskip=0pt\baselineskip=0pt
           \hbox{$<$}\hbox{$\sim$}}}}

\def\gsim{\mathrel{\lower2.5pt\vbox{\lineskip=0pt\baselineskip=0pt
           \hbox{$>$}\hbox{$\sim$}}}}

\begin{document}
\raggedright
\LARGE
Astro2020 APC White Paper \linebreak

\vspace{-4mm}
CMB-HD: An Ultra-Deep, High-Resolution Millimeter-Wave Survey Over Half the Sky\linebreak
\normalsize

\vspace{-3mm}
\noindent \textbf{Thematic Areas:} \linebreak
Primary Area: Cosmology and Fundamental Physics;~~Secondary Area: Galaxy Evolution \linebreak
Additional Areas: Planetary Systems, Compact Objects and Energetic Phenomena
\linebreak

\vspace{-3mm}  
\textbf{Primary Contact:}

Name: Neelima Sehgal
 \linebreak						
Institution:  Stony Brook University \& Flatiron Institute
 \linebreak
Email:  neelima.sehgal@stonybrook.edu;~~Phone:  631-632-8229
 \linebreak
 
\vspace{-3mm} 
\textbf{Proposing Team:} 
Simone Aiola$^{1}$,
Yashar Akrami$^{2}$,
Kaustuv Basu$^{3}$,
Michael Boylan-Kolchin$^{4}$,
Sean Bryan$^{5}$,
Sébastien Clesse$^{6,7}$,
Francis-Yan Cyr-Racine$^{8,9}$,
Luca Di Mascolo$^{10}$,
Simon Dicker$^{11}$,
Thomas Essinger-Hileman$^{12}$,
Simone Ferraro$^{13}$,
George M. Fuller$^{14}$,
Dongwon Han$^{15}$,
Mathew Hasselfield$^{16}$,
Gil Holder$^{17}$,
Bhuvnesh Jain$^{11}$,
Bradley Johnson$^{18}$,
Matthew Johnson$^{19,20}$,
Pamela Klaassen$^{21}$,
Mathew Madhavacheril$^{20}$,
Philip Mauskopf$^{5}$,
Daan Meerburg$^{22}$,
Joel Meyers$^{23}$,
Tony Mroczkowski$^{24}$,
Moritz Munchmeyer$^{20}$,
Sigurd Naess$^{1}$,
Daisuke Nagai$^{25}$,
Toshiya Namikawa$^{26}$,
Laura Newburgh$^{25}$,
\foreignlanguage{vietnamese}{Hồ Nam Nguyễn}$^{20,27}$,
Michael Niemack$^{28}$,
Benjamin D. Oppenheimer$^{29}$,
Elena Pierpaoli$^{30}$,
Emmanuel Schaan$^{13}$,
Neelima Sehgal$^{15,1}$,
An\v{z}e Slosar$^{31}$,
David Spergel$^{1}$,
Eric Switzer$^{12}$,
Alexander van Engelen$^{32,5}$,
Edward Wollack$^{12}$

\justify

\textbf{Abstract:}
A millimeter-wave survey over half the sky, that spans frequencies in the range of 30 to 350 GHz, and that is both an order of magnitude deeper and of higher-resolution than currently funded surveys would yield an enormous gain in understanding of both fundamental physics and astrophysics. By providing such a deep, high-resolution millimeter-wave survey (about 0.5 $\mu$K-arcmin noise and 15 arcsecond resolution at 150 GHz), CMB-HD will enable major advances. It will allow 1.)~the use of gravitational lensing of the primordial microwave background to map the distribution of matter on small scales ($k\sim10~h$Mpc$^{-1}$), which probes dark matter particle properties. It will also allow 2.)~measurements of the thermal and kinetic Sunyaev-Zel'dovich effects on small scales to map the gas density and gas pressure profiles of halos over a wide field, which probes galaxy evolution and cluster astrophysics. In addition, CMB-HD would allow us to cross critical thresholds in fundamental physics: 3.)~ruling out or detecting any new, light ($< 0.1$ eV), thermal particles, which could potentially be the dark matter, and 4.)~testing a wide class of multi-field models that could explain an epoch of inflation in the early Universe. Such a survey would also 5.)~monitor the transient sky by mapping the full observing region every few days, which opens a new window on gamma-ray bursts, novae, fast radio bursts, and variable active galactic nuclei. Moreover, CMB-HD would 6.)~provide a census of planets, dwarf planets, and asteroids in the outer Solar System, and 7.)~enable the detection of exo-Oort clouds around other solar systems, shedding light on planet formation. The combination of CMB-HD with contemporary ground and space-based experiments will also provide powerful synergies. CMB-HD will deliver this survey in 5
years of observing 20,000 square degrees, using two new 30-meter-class off-axis cross-Dragone telescopes to be located at Cerro Toco in the Atacama Desert. The telescopes will field about 2.4~million detectors (600,000 pixels) in total. The CMB-HD survey will be made publicly available, with usability and accessibility a priority.


\thispagestyle{empty}
\pagebreak
\setcounter{page}{1}

\section{Introduction} 

\vspace{-3mm}
An ultra-deep, high-resolution millimeter-wave survey over half the sky would answer many outstanding questions in both fundamental physics and astrophysics.  This survey would be both an order of magnitude deeper and of higher-resolution (about $0.5~\mu$K-arcmin and 15 arcseconds at 150 GHz) than currently funded surveys.  Two critical advances uniquely enabled by this survey are mapping over half the sky:~i)~the distribution of all matter on small scales using the gravitational lensing of the cosmic microwave background (CMB), and~ii)~the distribution of gas density and gas pressure on small scales in and around dark matter halos using the thermal and kinetic Sunyaev-Zel'dovich effects (tSZ and kSZ).  The combination of high-resolution and multiple frequency bands in the range of 30 to 350 GHz allows for critical separation of foregrounds from the CMB.  That plus the depth of the survey allows one to cross critical thresholds in fundamental physics:~i.)~ruling out or detecting any new, light, thermal particles, which could potentially be the dark matter, and~ii)~testing a wide class of multi-field models that could explain an epoch of inflation in the early Universe. CMB-HD would also open a new window on planetary studies and the transient sky.  A summary of the primary science questions motivating the CMB-HD survey are given below and discussed in more detail in the accompanying science white paper~\cite{Sehgal2019}. 

\vspace{-6mm} 
\begin{item}
\item $\bullet$ What is the distribution of matter on small scales? 
\item $\bullet$ What are the particle properties of dark matter?
\item $\bullet$ How did gas evolve in and around dark matter halos?
\item $\bullet$ How did galaxies form?
\item $\bullet$ Do new light particles exist that were in equilibrium with known particles in the early Universe?  
\item $\bullet$ Do primordial gravitational waves exist from an epoch of inflation?
\item $\bullet$ If inflation happened, did it arise from multiple or a single new field?
\item $\bullet$ What is the census of bodies in the outer Solar System?
\item $\bullet$ Do Oort clouds exist around other stars?
\item $\bullet$ What is the physics behind the various bright transient phenomena in the sky?
\end{item}

\vspace{-4mm}
\section{Key Science Goals and Objectives}

\vspace{-4mm}
\subsection{Fundamental Physics of the Universe} 

\vspace{-2mm}
There are fundamental questions about our Universe that beg answers. One such question is what is the nature of dark matter -- the invisible glue that is holding galaxies together, including our own.  It may be that dark matter only interacts gravitationally with the known particles, and in that case, astronomical observations may be the only pathway to understanding its properties.  Another question is whether we have a complete inventory of the light particles that were in equilibrium with the known particles in the early Universe.  Answering this question definitively is uniquely within reach of CMB-HD, and would probe physics right up to the Big Bang.  A final question is what is the origin of the Universe and all the `stuff' in it?  Did it start with a period of exponential inflationary expansion, and if so, how did that arise?  Addressing these questions are key science goals of CMB-HD.  Table~\ref{tab:goals} lists the approaches to addressing these questions, along with the science targets CMB-HD can achieve.  Below we explain each method and target in more detail.\\

\begin{table*}[t!]
\centering
\caption[Key Science]{Summary of CMB-HD key science goals in fundamental physics} \small
\vspace{0.5mm}
\begin{tabular}{ l c r }
\hline
\hline
Science & Parameter  & Sensitivity\\
& & \\
\hline
Dark Matter & S/N:~~Significance in Differentiating FDM/WDM from CDM\textsuperscript{a}  & S/N = 8 \\ 
New Light Species & $N_{\rm{eff}}$:~~Effective Number of Relativistic Species\textsuperscript{b} & $\sigma(N_{\rm{eff}}) = 0.014$ \\
Inflation & $f_{\rm{NL}}$:~~Primordial Non-Gaussianity\textsuperscript{c}  & $\sigma(f_{\rm{NL}}) = 0.26$ \\
Inflation &	$A_{\rm{lens}}$:~~Residual Lensing B-modes\textsuperscript{d} & $A_{\rm{lens}}=0.1$ \\
\hline
\hline
\end{tabular}
\begin{tablenotes}
\item \textsuperscript{a} 
Significance with which fuzzy/warm dark matter (FDM/WDM) models that can explain observational puzzles of small-scale structure, can be distinguished from cold dark matter.
\item \textsuperscript{b} 
{{Probing well below the critical threshold of $\sigma({N_{\rm{eff}}}) = 0.027$ would rule out or detect {\it{any}} new, light ($< 0.1$ eV) particle species in thermal equilibrium with standard model particles in the early Universe.}}
\item \textsuperscript{c} 
{{Achieving $\sigma(f_{\rm{NL}}) < 1$ can distinguish between single and multi-field inflation models.}}
\item \textsuperscript{d} 
{{Removing lensing B-modes is critical to detecting primordial gravitational waves from inflation; for example, $A_{\rm{lens}}\leq 0.1$ is necessary to achieve $\sigma(r)=5\times 10^{-4}$ from a ground-based experiment.}}
\end{tablenotes}
\label{tab:goals}
\vspace{-3mm}
\end{table*}

\vspace{0mm}
\noindent {\bf{Dark Matter:}}
Astronomical observations have provided compelling evidence for non-baryonic dark matter~\cite{Zwicky1937,Rubin1970,Ostriker1974,Fabricant1980,Bahcall1995,Clowe2006,Planck2018}. However, we have not been able to create or detect dark matter in manmade experiments to probe its properties directly.  If dark matter only interacts with the known standard-model Universe gravitationally, then it makes sense to explore the gravitational direction further to understand it.  An observational puzzle seems to exist regarding the distribution of matter on small-scales (scales below 10 kpc and masses below $10^9 M_{\odot}$ today); there seems to be less structure than the standard cold collisionless model of dark matter (CDM) would predict.  This may provide clues to the particle nature of dark matter, and, in fact, many well-motivated models of dark matter can explain this deviation from CDM~\cite{Colin2000,Bode2001,Boehm:2001hm,Boehm:2004th,Viel2005,Turner1983,Press1990,Sin1994,Hu2000,Goodman2000,Peebles_2000,Amendola2006,Schive2014,Marsh2016,Carlson1992,Spergel2000,Vogelsberger2012,Dvorkin:2013cea,Fry2015,Elbert2015,Kaplinghat2016,Kamada2017,Gluscevic2018,Boddy2018a,Li2018,Boddy2018b,Tulin2018,Khoury2016,Vogelsberger2016,Cyr-Racine:2013ab,Cyr-Racine:2015ihg,Schewtschenko:2014fca,Krall:2017xcw,Xu:2018efh}.

However, measurements of the small-scale matter distribution do not conclusively indicate a deviation from the CDM prediction.  This is because such measurements often infer the matter distribution through baryonic tracers~\cite{Koposov2015,Drlica-Wagner2015,Menci2017,Mesinger2005,deSouza2013,Moore1999,Johnston2016,Carlberg2009,Erkal2015,Bovy2014,Cen1994,Hernquist1996,Croft1999,Hui1999,Viel2013,Baur2016,Irsic2017}, and such tracers may not reliably map the dark matter~\cite{Sawala2016,Oman2015,Hui2017}. Gravitational lensing offers a powerful way to map the dark matter directly. While strong gravitational lensing is a promising method to find low-mass dark-matter halos~\cite{Dalal2002,Koopmans_2005,Vegetti:2009aa,Vegetti_2010_1,Vegetti_2010_2,Vegetti2012,Hezaveh:2012ai,Vegetti2014,Hezaveh2016a,Ritondale:2018cvp}, it does face the challenge of separating the complex and often unknown structure of the background source from the sought-after substructure signal; using strong lensing to measure the matter power spectrum, faces a similar challenge~\cite{Hezaveh2016b,Daylan:2017kfh,Bayer:2018vhy,DiazRivero2018,Chatterjee2018,Cyr-Racine:2018htu,Brennan2018,Rivero:2018bcd}.  A method that can evade this challenge is to {\it{measure the small-scale matter power spectrum from weak gravitational lensing using the CMB as a backlight}}~\cite{Nguyen2019}.  The CMB serves as a perfect backlight because i)~it has a known redshift, ii)~is behind every object, and iii)~is a known pure gradient on the small-scales of interest. Thus this method can provide a powerful complementary probe of dark-matter physics~\cite{Nguyen2019}. 

\begin{SCfigure}[1.4][t]
\centering
\includegraphics[width=0.49\textwidth, height=6.2cm]{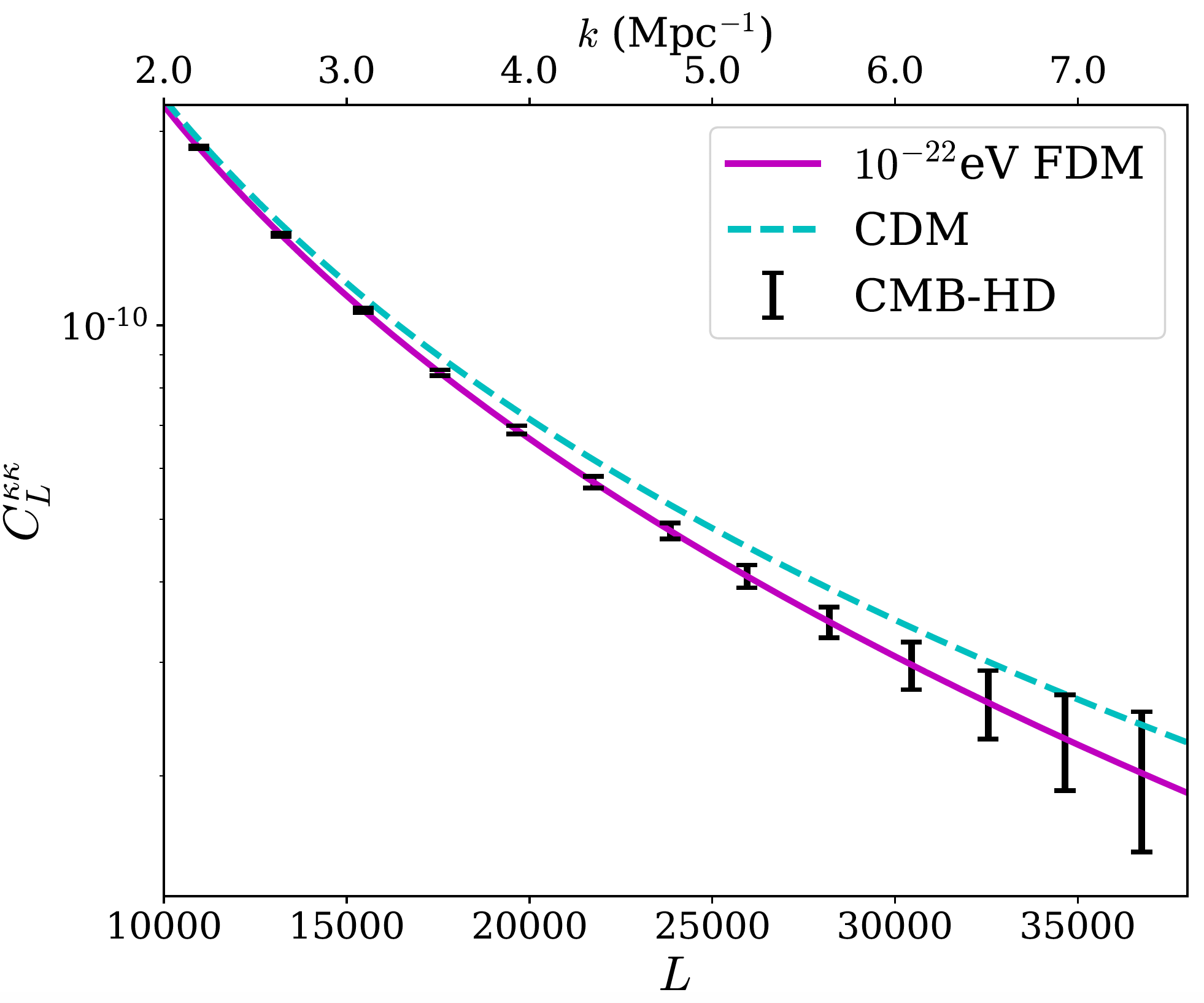}
\caption{CMB lensing power spectrum for an $m \sim 10^{-22}$ eV FDM model and a CDM model.  The error bars correspond to observations with $0.5 \mu$K-arcmin CMB noise in temperature and 15 arcsecond resolution over $50\%$ of the sky. Reionization kSZ has also been included as a foreground here. {\bf{Given these errors, one can distinguish between CDM and a suppression of structure below $10^9 M_{\odot}$ with a significance of about $8\sigma$.}}
}
\label{fig:dm}  
\vspace{-6mm}
\end{SCfigure}

We show in Figure~\ref{fig:dm} how one can distinguish between CDM and a model of dark matter that suppresses structure on small scales, by measuring the small-scale gravitational lensing of the CMB~\cite{Nguyen2019}. Table~\ref{tab:goals} indicates that one can distinguish between these two models at the $8\sigma$ level.  Newer lensing reconstruction techniques can potentially improve these constraints further~\cite{Horowitz2019,Hadzhiyska2019}. Extragalactic foregrounds are the main source of systematic effect in this measurement, and we discuss paths to mitigate this in~\cite{Nguyen2019, Sehgal2019} and in Section~\ref{sec:measureReq} below.  Baryonic processes can also move around the dark matter and change the shape of the small-scale matter power spectrum; however, they likely affect the shape of the spectrum in a way that differs from alternate dark matter models~\cite{Nguyen2019,vanDaalen2011,Brooks2013,Brooks2014,Natarajan2014,Schneider2018}.  Comparing the matter power spectrum of multiple hydrodynamic simulations, each with different baryonic prescriptions, suggests there may be a finite set of ways that baryons can change the shape of the spectrum, characterized by just a few free parameters~\cite{Schneider2018}.  Given that, one can use the shape of the power spectrum to distinguish between dark matter models and baryonic effects.  In any case, {\it{{\underline{this measurement would be a clean measurement of the matter power}}\\{\underline{spectrum on these scales, free of baryonic tracers.}}}} It would greatly limit the allowed models of dark matter and baryonic physics, shedding light on dark-matter properties and galaxy evolution. \\

\vspace{-2mm}
\noindent {\bf{New Light Species:}} The light from the CMB still has more to tell us about the inventory of particles that existed in the Universe right back to the Big Bang.  This is because any light particles (with masses less than 0.1 eV) that were in thermal equilibrium with the known standard-model particles, at any time in the early Universe, would have left an imprint on the CMB. Even one new species of particle as described above, would change the effective number of light particles, $N_{\rm{eff}}$, by as much as $\sigma({N_{\rm{eff}}}) = 0.027$ away from the standard model value (assuming no significant dilution by new states beyond the standard model particle content)~\cite{CMBS4SB}. 
If these light particles interact only very weakly, we may never see them in laboratory experiments, and astronomical measurements may provide the only avenue for their detection.
This is particularly important because many dark matter models predict new light thermal particles~\cite{Baumann:2016wac,Green:2017ybv}, and recent short-baseline neutrino experiments have found puzzling results possibly suggesting new neutrino species~\cite{Gariazzo2013,Gonzalez-Garcia2019}.

{\it{\underline{In Table~\ref{tab:goals}, we show that CMB-HD can achieve $\sigma({N_{\rm{eff}}}) = 0.014$, which would cross the critical}\\{\underline{threshold of 0.027.  This would potentially rule out or find evidence for new light thermal particles}}\\{\underline{with $95\%$ confidence level.}}}} 
We note that delensing the CMB spectra can modestly improve the $N_{\rm{eff}}$ constraints beyond this~\cite{Baumann:2015rya,Green:2016cjr}, and including the effect of Rayleigh scattering can potentially have a significant impact (e.g.~25\% improvement with the inclusion of the 280/350 GHz frequency channels)~\cite{Lewis2013,Alipour:2014dza}.  By combining the CMB-HD primordial CMB measurements with the CMB-HD measurement of the small-scale matter power spectrum discussed above, we can also potentially gain a factor of two improvement on $\sigma({N_{\rm{eff}}})$, subject to improvements in modelling the nonlinear matter power up to $k\sim0.4~h$Mpc$^{-1}$~\cite{Baumann2018}. Removal of foregrounds is important for achieving these constraints, and we discuss paths to do so in~\cite{Sehgal2019} and in Section~\ref{sec:measureReq} below; we note here that the high-angular resolution and multiple frequency channels of CMB-HD, spanning 30 to 350 GHz, help make the necessary foreground cleaning possible. The combination of above improvements to the nominal forecast in Table~\ref{tab:goals}, could potentially rule out or find evidence for new light thermal particles at the $99\%$ confidence level.\\

\noindent {\bf{Inflation:}} A fundamental question is what happened right at the Big Bang, and how did the structure we see (galaxies, planets, us) arise out of the homogeneous soup of the newborn Universe.  Inflation is one compelling idea in which the Universe underwent a period of superluminal expansion a fraction of a second after the Big Bang; this expansion stretched out microscopic quantum fluctuations and froze them into the fabric of spacetime as macroscopic density fluctuations, seeding the structure we see today.  However, we have no direct evidence that inflation actually occurred.  

One prediction of inflation is that fluctuations of the quantum particle carrying gravity created gravitational waves that later got imprinted in the CMB. We can detect this imprint by looking for a well-defined signal in the polarized CMB light, in particular in the large-scale CMB B-mode fluctuations.  The amplitude of this primordial B-mode fluctuation is given by the parameter $r$, and a number of well-motivated models predict $r \sim 3 \times 10^{-3}$; thus $\sigma(r)<5 \times 10^{-4}$ has become a desired target to definitively rule them out~\cite{CMBS4SB}.  However, to reach this target using a ground-based experiment requires removing the CMB B-mode fluctuations from gravitational lensing.  In particular, for this $\sigma(r)$ target, one must remove 90\% of the lensing B-mode power, leaving only 10\% remaining. {\it{\underline{Table~\ref{tab:goals} shows that CMB-HD is easily able to reach this $A_{\rm{lens}}=0.1$ target.}}}

In addition, inflation predicts the primoridal CMB has small non-Gaussian fluctuations that can be characterized by a parameter called $f_{\rm{NL}}$~\cite{Meerburg2019}.  Reaching a target of $\sigma(f_{\rm{NL}}) < 1$ would rule out a wide class of multi-field inflation models, shedding light on how inflation happened~\cite{Alvarez:2014vva,Smith2018,Munchmeyer2018,Deutsch2018,Contreras2019,Cayuso2018}.  By combining the kSZ signal from CMB-HD, which is discussed in more detail in Section~\ref{sec:astro}, with an overlapping galaxy survey such as LSST, {\it{\underline{Table~\ref{tab:goals} shows that one can cross this critical threshold,}\\{\underline{achieving $\sigma(f_{\rm{NL}}) = 0.26$.}}}}  This cross-correlation could also resolve the physical nature of several statistical anomalies in the primary CMB~\cite{Cayuso2019} that may suggest new physics during inflation (see Ref.~\cite{Schwarz2015} for a review), and provide constraints on the state of the Universe before inflation~\cite{Zhang2015}.

\vspace{-3mm}
\subsection{Astrophysics}\label{sec:astro} 

\vspace{-2mm}
\noindent {\bf{Evolution of Gas:}} A fundamental question in astrophysics is: how did galaxies form? A critical advance in answering this question would be measuring the density, pressure, temperature, and velocity of the gas in and around dark matter halos out to $z\sim 2$ and with masses below $10^{12}$ M$_\odot$.  This is because the gas in these halos reflects the impact of feedback processes and mergers, and it serves as a reservoir enabling star formation.  However, to date we have not had such measurements over a statistically significant sample. 

\begin{figure}[t!]
\centering
\vspace{-5mm}
\includegraphics[width=0.8\textwidth,height=7cm]{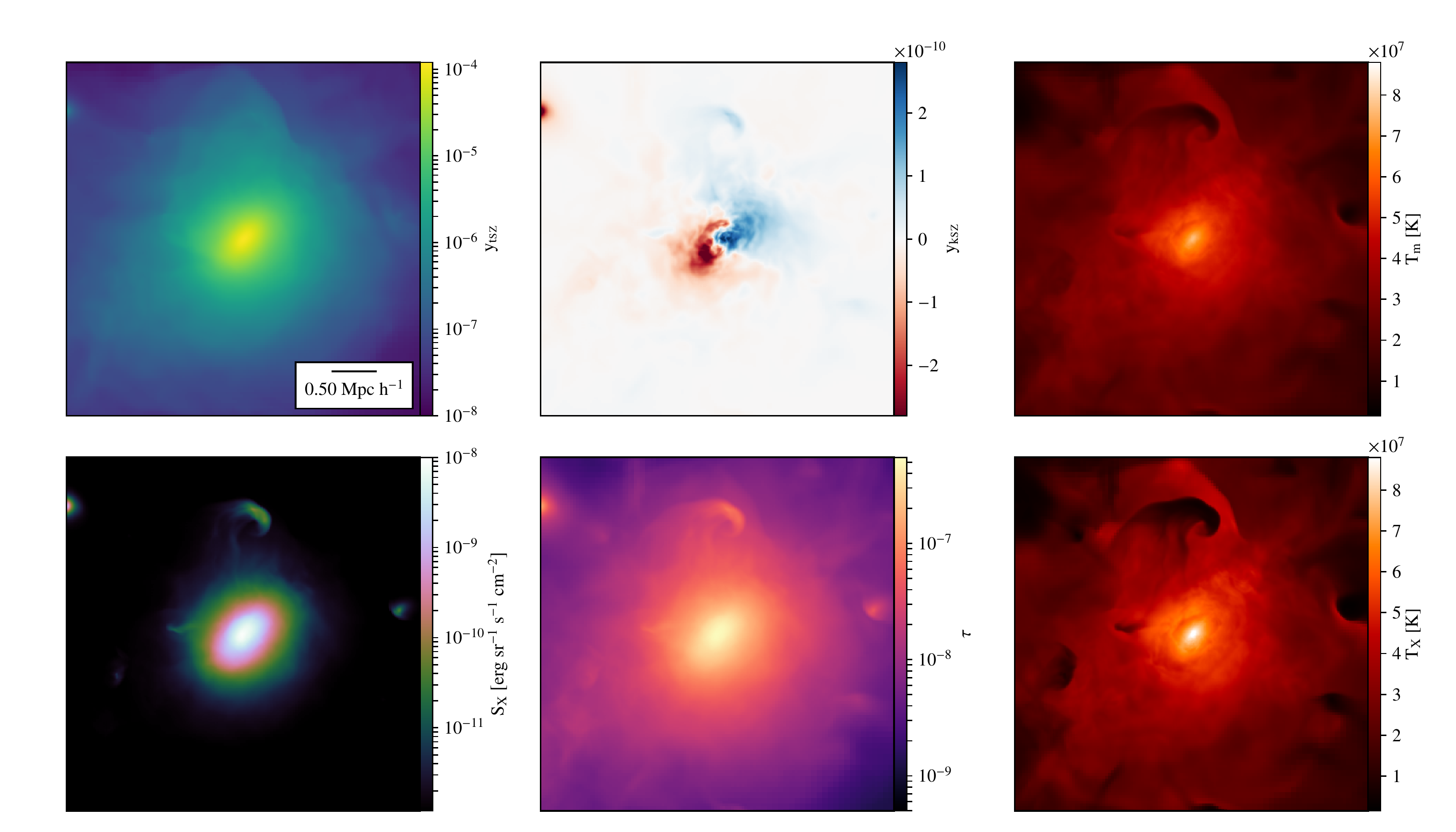}
\vspace{-6mm}
\caption{SZ signals of a merging galaxy cluster extracted from the {\it Omega500} cosmological hydrodynamical simulation \cite{Nelson2014a}; for comparison some X-ray quantities are shown. Here is the tSZ signal (top-left), kSZ signal (top-middle), projected mass-weighted temperature (top-right), X-ray surface brightness (bottom-left), electron opacity (bottom-middle), and projected X-ray temperature (bottom-right).
{\bf{The combination of tSZ and kSZ measurements from CMB-HD allows one to separate the pressure, density, temperature, and velocity profiles of the gas}}~\cite{Knox2004,Sehgal2005,Battaglia2017}.}
\label{fig:SZ}  
\vspace{-4mm}
\end{figure}

CMB-HD would open this new window by measuring the tSZ and kSZ effects with high-resolution and low-noise over half the sky (for a recent SZ review see~\cite{Mroczkowski2019}).  The tSZ signal is a measure of the thermal pressure of ionized gas in and around dark matter halos~\cite{Sunyaev1970}.  The kSZ signal measures the gas momentum density~\cite{Sunyaev1972}. The combination of both tSZ and kSZ measurements, with low-noise and high-resolution across CMB-HD frequencies,
would {\it{\underline{allow CMB-HD to}\\{\underline{separately measure the gas density, pressure, temperature, and velocity profiles of the gas, as a}}\\{\underline{function of halo mass and redshift}}}}~\cite{Knox2004,Sehgal2005,Battaglia2017}.  As seen in Figure~\ref{fig:SZ}, this would probe thermal, non-thermal, and non-equilibrium processes associated with cosmic accretion, merger dynamics, and energy feedback from stars and supermassive black holes, and their impact on the gas~\cite{Nagai2011,Nelson2014b,Lau2015,Avestruz2015,Basu2016}.

Measuring the tSZ effect is also an effective way to {\it{find}} new galaxy clusters and groups, as has been well-demonstrated over the past decade~\cite{Marriage2011,Vanderlinde2010,Planck2014}. 
Millimeter-wave SZ measurements have an advantage over X-ray measurements in probing gas physics because the SZ signals are proportional to the gas density (not density squared) and the brightness of the signals are redshift independent; this makes SZ measurements a powerful probe of the gas in the outskirts of galaxy clusters, and in low-mass and high-redshift halos.  CMB-HD will push halo-finding to lower masses and higher redshifts, allowing direct imaging of systems where X-ray observations would require prohibitively long integration times. {\it{\underline{By stacking the tSZ-detected halos, CMB-HD enables probing the}\\{\underline{gas physics in and around halos out to $z\sim 2$ and with masses below $10^{12}$ M$_\odot$.}}}}  The circumgalactic reservoirs of $10^{12}$ M$_\odot$ (Milky-Way-mass) halos are predicted by multiple simulations, such as EAGLE and Illustris-TNG, to be intimately linked to the appearance of, and activity within, the galaxy~\cite{schaye2015,nelson2018a,davies2019,pillepich2018}.  These and other simulations find that galactic star-formation rates, colors, and morphologies are inextricably linked not only to the mass in the circumgalactic medium, but also to the location of baryons ejected beyond $R_{200,c}$, which can be uniquely constrained by CMB-HD. Thus the science gain of such measurements is a more complete understanding of galaxy cluster astrophysics, the physics of the intergalactic and circumgalactic medium, and galaxy evolution.\\

\vspace{-1mm}
\noindent {\bf{Planetary Studies:}} The complete inventory of planets, dwarf planets, and asteroids in our own Solar System remains an open question.  CMB-HD can open a new discovery space in our outer Solar System by detecting undiscovered Solar System bodies via their thermal flux and parallactic motion~\cite{Cowan2016}. Objects close to the Sun are normally detected via optical observations, which are sensitive to the bodies' reflected light from the Sun.  However, objects also have internal heat that is emitted at millimeter wavelengths.  Since the flux from reflected light falls faster with distance than directly sourced emission, CMB-HD has an advantage over optical surveys in finding objects in the far Solar System~\cite{Baxter2018a}. In particular, {\it{\underline{CMB-HD could detect dwarf-size planets hundreds of}\\{\underline{AU from the Sun, and Earth-sized planets thousands of AU from the Sun.}}}}  In combination with  optical measurements, CMB-HD would allow large population studies of the sizes and albedos of these objects~\cite{Gerdes:2017}. In addition, whether our Solar System, and exo-solar systems in general, possess an Oort cloud is still unknown.  {\it{\underline{The low-noise and high-resolution of CMB-HD would enable}\\{\underline{the detection of exo-Oort clouds around other stars, opening a new window on planetary studies.}}}}\\

\noindent {\bf{Transient Sky:}}  CMB-HD will survey half the sky with roughly daily cadence.  The daily maps will have a noise sensitivity of about 1 mJy.  Thus CMB-HD will have excellent sensitivity to bright time-variable and transient sources in the sky~\cite{Holder2019}.  For example, CMB-HD will detect of order 100 on-axis long gamma-ray bursts (LGRBs)~\cite{Metzger2015}.  These LGRBs often are bright in the millimeter weeks before they peak in the radio, and the additional frequency coverage can help characterize their shock behavior~\cite{Laskar2018}.  Blazars, jet-dominated active galactic nuclei, are bright in the millimeter and vary significantly at about 100 GHz on week timescales~\cite{Chen2013,Madejski2016,Blandford2018}.  Novae, repeating thermonuclear explosions of accreting white dwarfs, are also millimeter-bright, with expected rates of about ten per year in our Galaxy.  CMB-HD will provide unique insight into the geometry of blazars and novae by providing polarization flux and variability information for these systems~\cite{Abdo2010,Young2010}. In addition, some neutron star mergers may be seen and localized first in the millimeter band~\cite{Fong2015}.  The CMB-HD survey would thus provide useful follow-up of LIGO/Virgo triggers, and could provide blind detections of these events~\cite{Holder2019}. {\it{\underline{The intent of the CMB-HD project is to provide to the}}\\{\underline{astronomy community weekly maps of the CMB-HD survey footprint, filtered to keep only small}}\\{\underline{scales, and with a reference map subtracted to make variability apparent.}}}   

\vspace{-3mm}
\section{Technical Overview: Measurement Requirements}\label{sec:measureReq} 
\vspace{-3mm}
The measurement requirements are driven by the dark matter and new light species science targets, since those targets set the most stringent requirements.  All the other science targets benefit from and prefer these same measurement requirements, since there are no science targets that pull the requirements in opposing directions.  This results in a fortunate confluence between the science targets presented above and the measurement requirements that can achieve them. \\

\vspace{-2mm}
\noindent {\bf{Sky Area:}} Both the dark matter and new light species science targets prefer the widest sky area achievable from the ground, given fixed observing time~\cite{Sehgal2019,CMBS4SB}.  For the dark matter science case, the inclusion of the kSZ foregrounds is what makes wider sky areas preferable over smaller ones~\cite{Sehgal2019}.  In addition, the non-Gaussianity inflation science target, searches for planets and dwarf planets, and probing the transient sky all benefit from the largest sky areas possible.  In practice, this is about {\it{\underline{50\% of the sky}}}, achievable from the demonstrated site of the Atacama Desert in Chile. \\ 

\begin{SCfigure}[1.4][t]
\centering
\includegraphics[width=0.55\textwidth,height=6.5cm]{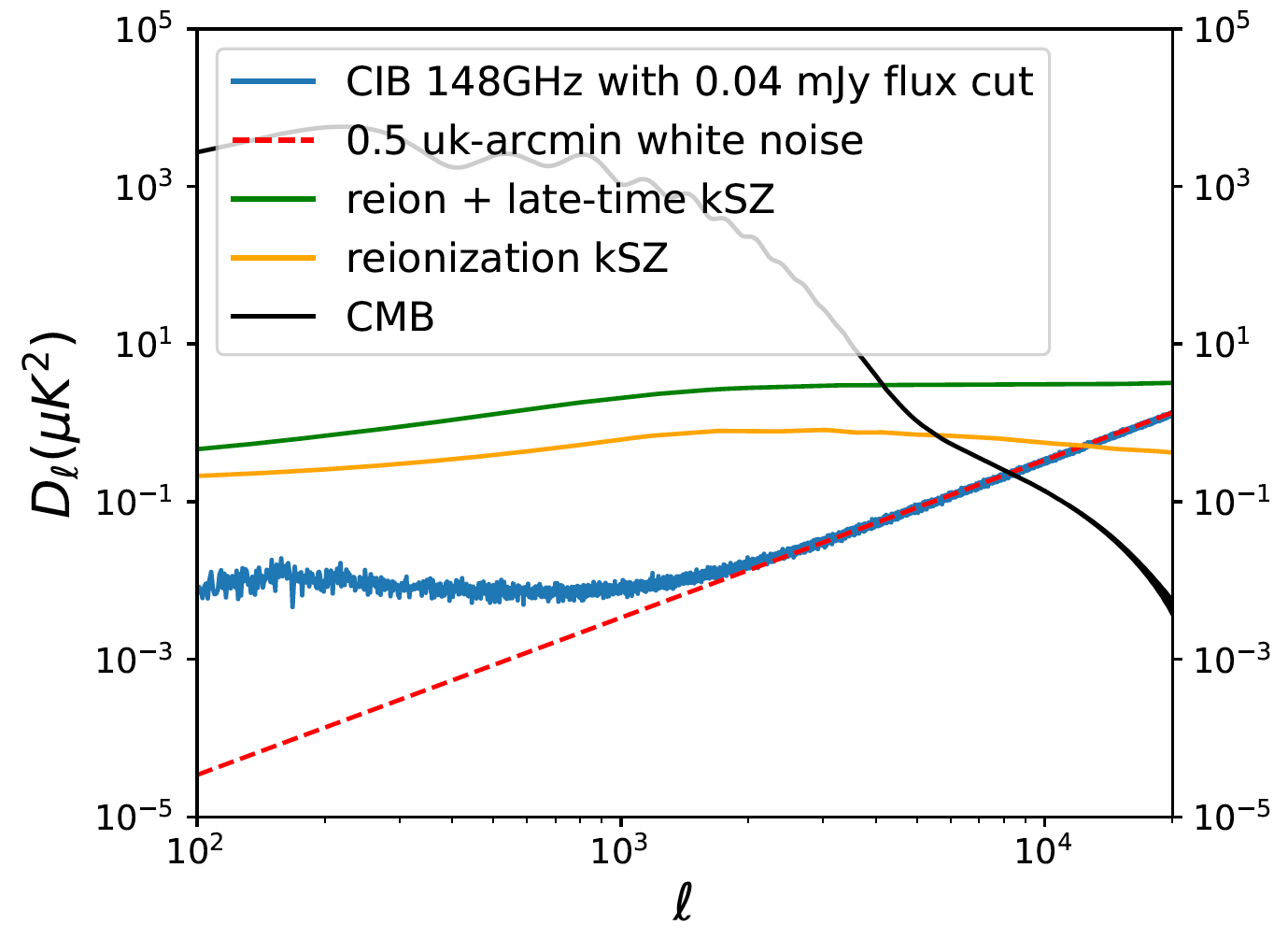}
\caption{Shown are the CMB temperature power spectrum (black solid) and relevant foregrounds at 150 GHz.  The foregrounds are the kSZ effect from the epoch of reionization (orange), reionization kSZ plus the late-time kSZ effect (green), and the CIB (after removing sources above a flux of 0.04 mJy).  The CIB flux cut, enabled by frequency channels between 100 and 350 GHz and the 30-meter dish, brings the CIB to the level of 0.5~$\mu$K-arcmin (dashed red) on small scales.}
\label{fig:foregrounds}  
\vspace{-5mm}
\end{SCfigure}

\vspace{-1mm}
\noindent {\bf{Resolution:}} The resolution is set by the dark matter science target of measuring the matter power spectrum on comoving scales of $k\sim10 h$Mpc$^{-1}$ (these scales collapsed to form masses below $10^9 M_\odot$ today).  Since CMB lensing is most sensitive to structures at $z\sim 2$ (comoving distance away of 5000 Mpc), we need to measure a maximum angular scale of $\ell \sim k X \sim 35,000$. {\it{\underline{This gives a required resolution of about 15 arcseconds at 150 GHz}}} ($\ell = \pi$/radians), translating to a 30-meter telescope.  In addition, foreground cleaning will be essential, and the most dominant foreground will be extragalactic star-forming galaxies, i.e.~the Cosmic Infrared Background (CIB).  Figure~\ref{fig:foregrounds} shows that removing the sources detected at $3\sigma$ at 280 GHz with a 30-meter dish (flux cut of about 0.2 mJy~\cite{Lagache2018}) from the 150 GHz maps (flux cut of 0.04 mJy~\cite{Sehgal2010}) lowers the CIB power at 150 GHz to below $0.5 \mu$K-arcmin.  
A resolution of 15 arcsecond is also needed to measure the profiles of the gas in halos and to separate extragalactic radio and star-forming galaxies from the gas signal.  To obtain a census of objects in the outer Solar System, we note that the parallactic motion of objects 10,000 AU away from the Sun is about 40 arcseconds in a year, also requiring this minimum resolution to detect the motion across a few resolution elements.  \\

\vspace{-8mm}
\noindent {\bf{Sensitivity:}} The sensitivity is driven by the dark matter science target. In order to measure at the $5\sigma$-level a deviation from the CDM-only prediction of the matter power spectrum that matches claimed observations of suppressed structure, and conservatively assuming one cannot remove any of the kSZ foregrounds shown in Figure~\ref{fig:foregrounds}, requires keeping the noise subdominant to the kSZ foregrounds out to $\ell \sim 25,000$.  {\it{\underline{This requires 0.5~$\mu$K-arcmin noise in temperature (0.7~$\mu$K-arcmin in}}\\{\underline{polarization) in a combined 90/150 GHz channel.}}} Assuming removal of the late-time kSZ component with an overlapping galaxy survey like LSST, an $8\sigma$ deviation from the CDM-only prediction can be seen (see Table~\ref{tab:goals} and Figure~\ref{fig:dm}).  
Conveniently, this sensitivity level also allows one to cross critical thresholds, achieving $\sigma({N_{\rm{eff}}}) = 0.014$ and $\sigma(f_{\rm{NL}}) = 0.26$, as well as to delens 50\% of the sky to $A_{\rm{lens}}=0.1$, a level necessary for reaching $\sigma(r)<5 \times 10^{-4}$ from a ground-based experiment. \\

\vspace{-2mm}
\noindent {\bf{Largest Angular Scale:}} By the time CMB-HD has first light, data from the Simons Observatory (SO)~\cite{SO2019}, which will have first light in 2021, will already exist and be public.  SO will measure temperature and E-mode maps over half the sky to the sample variance limit in the multipole range of 30 to 3000 for temperature and 30 to 2000 for E-modes.  SO will also measure these scales at six different frequencies spanning 30 to 280 GHz.  Thus there is no need for CMB-HD to reimage these modes.  Therefore the largest angular scale CMB-HD needs to measure is driven by measuring B-modes, and in particular being able to remove lensing B-modes to the level of $A_{\rm{lens}}=0.1$.  {\it{\underline{This requires the largest scale CMB-HD needs to image to be about 10 arcminutes ($\ell \sim 1000)$}}}. \\

\vspace{-2mm}
\noindent {\bf{Frequency Coverage:}} The frequency coverage is driven by needing most of the sensitivity in the frequency window that is most free from extragalactic foregrounds, namely 90 to 150 GHz.  Modern detectors can observe at two frequencies simultaneously~\cite{Henderson2016,Benson2014,OBrient2013}, so we assume we can split closely spaced frequency bands, further helping to remove frequency-dependent foregrounds.  We also require frequency coverage at 30/40 GHz to remove emission from radio galaxies and at 220/280 GHz to remove emission from dusty galaxies and to cover the null frequency of the tSZ signal (at 220 GHz).  Foreground optimization studies done for SO have found optimal ratios of noise levels given their six frequency channels, which if extrapolated to {\it{\underline{CMB-HD would require}}\\ {\underline{noise levels in temperature maps of 6.5/3.4, 0.73/0.79, and 2/4.6 $\mu$K-arcmin for the 30/40, 90/150,}}\\ {\underline{and 220/280 GHz channels respectively.}}} In addition, we require a 280/350 GHz channel in order to better exploit the Rayleigh scattering effect for improving the $N_{\rm{eff}}$ constraint~\cite{Lewis2013,Alipour:2014dza}, and in order to remove the main foreground (the CIB).  To clean the CIB to the level shown in Figure~\ref{fig:foregrounds} by removing $3\sigma$ sources
{\it{\underline{requires a 280/350 GHz channel with about 3.25~$\mu$K-arcmin combined noise}}}. 

\vspace{-4mm}
\section{Technical Overview: Instrument Requirements} 

\vspace{-3mm}
Given the measurement requirements above needed to achieve the science targets, the following instrument specifications below are required. \\ 

\vspace{-3mm}
\noindent {\bf{Site:}} The required sky area and sensitivity make {\it{\underline{Cerro Toco in the Atacama Desert the best site for}\\{\underline{CMB-HD}}}}.  An instrument at this site can observe the required 50\% of the sky; in contrast, an instrument at the South Pole can see a maximum of 6\% of the sky.  The sensitivity requirement of 0.5~$\mu$K-arcmin also requires locating CMB-HD at a high, dry site with low precipitable water vapor to minimize the total number of detectors needed.  No site within the U.S. has a suitable atmosphere.  While a higher site than Cerro Toco, such as Cerro Chajnantor in the Atacama Desert, might reduce the detector count further, that may not outweigh the increased cost of the higher site.\\

\vspace{-3mm}
\noindent {\bf{Detectors:}} To reach the required sensitivity levels of 6.5/3.4, 0.73/0.79, and 2/4.6 $\mu$K-arcmin for the 30/40, 90/150, and 220/280 GHz channels respectively, requires scaling down the SO goal noise levels by a factor of 8~\cite{SO2019}.  SO, which is at the same site as CMB-HD, requires 30,000 detectors to achieve its sensitivity levels~\cite{SO2019}.  Since the noise level scales as sqrt$(N_{\rm{det}})$, {\it{\underline{CMB-HD}\\{\underline{requires 1.9 million detectors to reach a factor of 8 lower noise than SO across all frequencies. An}\\{\underline{extra 280/350 GHz channel with 3.25~$\mu$K-arcmin noise increases the detector count to 2.4~million.}}}}} \\This assumes a 5-year survey and 20\% observing efficiency, as also assumed by SO.\\

\vspace{-3mm}
\noindent {\bf{Telescope Dish Size:}} To achieve the required resolution of 15 arcseconds at about 100 GHz requires {\it{\underline{a telescope dish size of about 30-meters}}}.  The foreground cleaning discussed above also necessitates a 30-meter dish for the frequencies above 100 GHz, out to at least 350 GHz.

\vspace{-5mm}
\section{Technical Overview: Instrument Design} 
\vspace{-3mm}
\subsection{Baseline Instrument}
\vspace{-2mm}
The CMB-HD telescope cameras will hold 600,000 pixels.  Each pixel will have two frequency bands and two polarizations for a total of 2.4~million detectors.  We assume in the baseline design horn-fed TES detectors, however, MKIDS may also be a viable detector technology, which could reduce the cost significantly.  There will be four frequency band pairs:~30/40, 90/150, 220/280, 280/350 GHz.  The distribution of detectors per frequency for the first three band pairs will follow the ratios adopted by SO, and achieve the noise levels given above.

\begin{figure}
\centering
\includegraphics[width=0.95\textwidth]{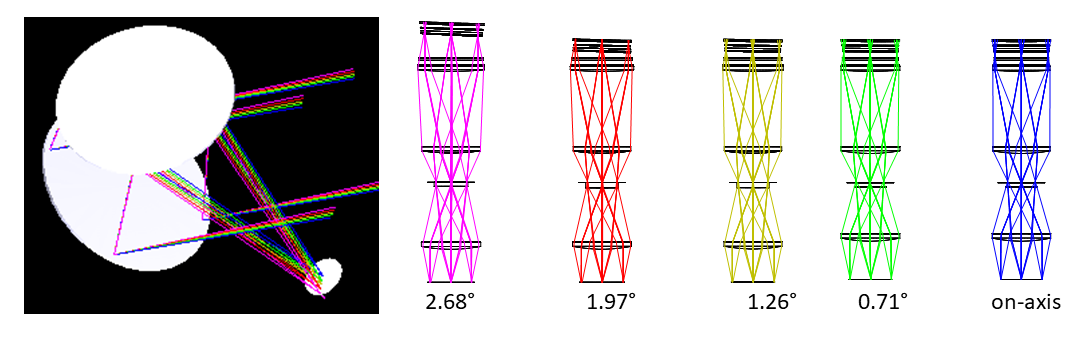}
\vspace{-7mm}
\caption{A possible design for the optics would be a crossed Dragone design similar to that chosen for SO and CCAT-prime.  Although challenging to build, a 30-meter version of this particular design offers a diffraction limited field of view out to $r=0.8$~degrees at 150~GHz at the secondary focus, and, with correcting optics such as those shown on the right, diffraction limited beams can be achieved out past a radius of 2.68 degrees at 150~GHz and even further at lower frequencies.  These particular designs are limited by the largest silicon optics currently available (45~cm) and could be grouped together in multiple cryostats in order to ensure less down time and greater flexibility.}
\label{fig:optics}  
\vspace{-3mm}
\end{figure}

For systematic control, the baseline design is an off-axis telescope.  We use a crossed Dragone design because it has a larger field-of-view (fov) than Gregorian or Cassegrain telescopes.  Although the crossed Dragone design has a far larger secondary mirror, as shown in Figure~\ref{fig:optics}, with the use of correcting cold optics extremely large fovs are possible. Simple calculations indicate that with efficient focal plane use ($>50$\%) it could be possible to fit all the detectors in one telescope.  However, given the number of receivers required, the baseline design consists of two telescopes. 

For a baseline receiver we adopt a design similar to that for CCAT-prime and SO.  Multiple sets of cold silicon lenses re-image the telescope focal plane while adding a 1K lyot stop, baffles for control of stray light, and cold blocking and bandpass filters.  For SO and CCAT-prime these are housed in a single cryostat. However, for a telescope the size CMB-HD it makes sense to group them into a number of cryostats; a single cryostat many meters across is hard to build and impractical to transport and maintain. In addition, recent advances in low loss silicon~\cite{Chesmore2018} could allow a warm first lens, and in that case the packing density of tubes could become far greater~\cite{Niemack2016}, allowing for smaller cheaper cryostats.  

\subsection{Cost Estimates and Technology Drivers} 
\vspace{-2mm}
The CMB-HD instrument and site cost is shown in Table~\ref{tab:cost} with 2019 estimates. Most of the cost is in fabricating the detectors.  For detector cost comparison, CMB-HD will have a factor of 4.8 more detectors than the proposed CMB-S4 experiment (2400k for \$960 million versus 500k for \$200 million).  However, there are two avenues that could potentially reduce this cost.  One is that over the next several years, as upcoming experiments such as SO require tens of thousands of detectors, the mass production of detectors may drop the cost.  The other is that MKIDS, a detector technology alternate to TES devices, are currently being tested on-sky at millimeter wavelengths~\cite{Austermann2018}. If MKID detector technology matches in practice to its promise, then the detector cost may drop by a factor of a few.  Two 30-meter off-axis, crossed Dragone telescopes are costed at \$400 million.
{\it{\underline{The unprecedented sensitivity and resolution of CMB-HD opens up vast uncharted territory in the}}\\  {\underline{millimeter-wave sky, and is a clear leap beyond currently funded ground-based CMB experiments.}}}

\vspace{-4mm}
\section{Data Analysis and Products}
\vspace{-3mm}
A major effort of the collaboration team will be the creation of data analysis software pipelines as well as simulations to verify these pipelines.  The core project deliverables, after a modest proprietary period, will be well-characterized maps at each frequency from which the broader science community can derive a range of science results.  Other project deliverables will be the power spectra, covariance matrices, likelihood functions, and Monte Carlo simulation sets.  The project also intends to provide the astronomy community with weekly maps of the CMB-HD survey footprint, filtered to keep only small scales, and with a reference map subtracted to make variability apparent.

\vspace{-5mm}
\begin{table*}[b]
\vspace{-5mm}
\centering
\begin{tabular}{| l r |}
\hline
\$960 million &  2.4 million detectors (600k pixels) and their readout \\ 
\$400 million  & Two 30-meter off-axis crossed Dragone telescopes \\
\$100 million & Project management \\
\$~50 million   & Cerro Toco site in the Atacama Desert \\
\hline
\$1.51 billion & Total cost\\
\hline
\end{tabular}
\vspace{-2mm}
\caption{Cost of CMB-HD Instrumentation and Construction}
\label{tab:cost}
\vspace{-5mm}
\end{table*}

\newpage
\vspace{-4mm}
\section{CMB-HD Project} 
\vspace{-4mm}
\begin{center}
\linespread{0.9}
\noindent \large{{\it{{Vision statement: One day, we will know how the Universe began, what is in it, and where it is going.}}}}
\end{center}

\vspace{-3mm}
\begin{center}
\linespread{0.9}
\noindent \large{{\it{{Mission statement: To further humanity's understanding of the Universe.  To build a culture of caring, connection, responsibility, and trust within the project team. To share the excitement of science with the public.}}}}
\end{center}

\vspace{-8mm}
\subsection{Project Philosophy}
\vspace{-2mm}
With endeavors as ambitious as CMB-HD, which can be a source of national pride and which can radically alter our understanding of the Universe and our place within it, it is important to articulate the founding intention and develop the project in alignment with that.  As stated in the vision statement above, the aspirational goal of the CMB-HD project is to answer questions asked by every human, namely: ``What is the Universe we live in, how did it start, and how will it progress?''  To accomplish this, the CMB-HD project's mission is to further our knowledge of the Universe on many fronts.  In addition, the process of getting there is just as important.  The CMB-HD philiosophy is that engaged teams are more productive and creative, and that engaged teams are created when leaders have fostered a culture of caring, connection, responsibility, and trust. This philosophy will drive the collaboration structure and all project interactions. The CMB-HD team also recognizes that thirst for knowledge about our Universe is universal, and that it is a privilege to be on the front lines of exploration.  Therefore, part of the mission of the CMB-HD project is to communicate and share the excitement of scientific discovery with the broader public.      

\vspace{-4mm}
\subsection{Organization, Partnerships, and Current Status}
\vspace{-2mm}
The CMB-HD collaboration was recently formed and is open to the entire U.S. science community.  International partners are welcome.  We recognize the CMB-S4 Membership Policy and the Simons Observatory Publication Policy as good starting documents to build from (both of which were co-drafted by the CMB-HD primary contact).  We will draft the governance policy in full transparency and consultation with the CMB-HD collaboration, building on the CMB-community-led work that informed the CMB-S4 Governance Policy, and being open to aspects that can be improved.   

\vspace{-5mm}
\section{Synergies with Other Millimeter-Wave Proposals}
\vspace{-4mm}
We recognize that there exists common ground with both the CMB-S4 and AtLAST project proposals~\cite{CMBS4SB, Bertoldi2018}, and potential synergistic opportunities.  
{\it{\underline{We also note that CMB-HD can achieve many}}\\  {\underline{of the science goals of the original CCAT 25-meter proposal~\cite{CCAT}, which was highly ranked in the}}\\  {\underline{Astro2010 Decadal}}} but not pursued.  Although CMB-HD does not extend as high as CCAT into the submillimter, with 350 GHz being the highest channel, CMB-HD's much higher sensitivity as compared to CCAT uniquely enables the science described above.

\vspace{-6mm}
\section{Summary} 
\vspace{-3mm}
CMB-HD is an ambitious leap beyond previous and upcoming ground-based millimeter-wave experiments. It will allow us to cross critical measurement thresholds and definitively answer pressing questions in astrophysics and fundamental physics of the Universe.  The CMB-HD project also recognizes the productivity benefit of an engaged collaboration, and is committed to fostering a culture that enables this.  The CMB-HD survey will be made publicly available, and the project will prioritize usability and accessibility of the data by the broader scientific community.

\clearpage
\section*{Affiliations}

\noindent $^{1}$ Flatiron Institute \\
$^{2}$ École Normale Supérieure, Paris, France \\
$^{3}$ University of Bonn \\
$^{4}$ The University of Texas at Austin \\
$^{5}$ Arizona State University \\
$^{6}$ University of Louvain, Belgium \\
$^{7}$ University of Namur, Belgium \\
$^{8}$ Harvard University \\
$^{9}$ University of New Mexico \\
$^{10}$ Max Planck Institute for Astrophysics \\
$^{11}$ University of Pennsylvania \\
$^{12}$ NASA/Goddard Space Flight Center\\
$^{13}$ Lawrence Berkeley National Laboratory \\
$^{14}$ University of California, San Diego \\
$^{15}$ Stony Brook University \\
$^{16}$ Pennsylvania State University \\
$^{17}$ University of Illinois at Urbana-Champaign \\
$^{18}$ Columbia University \\
$^{19}$ York University \\
$^{20}$ Perimeter Institute \\
$^{21}$ UK Astronomy Technology Centre \\
$^{22}$ University of Cambridge \\
$^{23}$ Southern Methodist University \\
$^{24}$ European Southern Observatory \\
$^{25}$ Yale University \\
$^{26}$ Cambridge University \\
$^{27}$ University of California, Berkeley \\
$^{28}$ Cornell University \\
$^{29}$ University of Colorado \\
$^{30}$ University of Southern California \\
$^{31}$ Brookhaven National Laboratory \\
$^{32}$ Canadian Institute for Theoretical Astrophysics


\newpage

\bibliographystyle{unsrturltrunc6}
\bibliography{main.bbl}

\end{document}